\documentclass{spie}
\usepackage{amsmath,amsfonts,amssymb}
\usepackage{graphicx}
\usepackage[colorlinks=true, allcolors=blue]{hyperref}

\let\h\^

\def\[#1\]{\begin{align}#1\end{align}}
\def\=#1{\bar{\bar{\mathrm{#1}}}}
\def\##1{\mathfrak{#1}}
\def\^#1{\mathbf{\hat #1}}
\def\_#1{\mathbf{#1}}
\def\o{\omega}
\def\O{\Omega}
\def\.{\cdot}
\def\x{\times}
\def\xx{{\!\!\tiny\begin{array}{c}\times\\\times\end{array}\!\!}}
\def\ox{\mathbin{\raisebox{1pt}{$\scriptstyle\otimes$}}}
\def\E{\varepsilon}
\def\M{\mu}
\def\l#1{\label{eq:#1}}
\def\r#1{(\ref{eq:#1})}
\def\d{\partial}
\def\D{\nabla}
\def\Dt{\nabla_{\rm t}}

\def\matr#1{\begin{pmatrix}#1\end{pmatrix}}

\def\Im{\mathrm{Im}}
\def\bra{\langle}
\def\ket{\rangle}
\def\eg{{\it e.g.}}
\title{Dynamics of slowly varying fields in bianisotropic media}

\author[\relax]{Stanislav I. Maslovski}
\affil[\relax]{Instituto de Telecomunica\c{c}\~{o}es, DEEC
  FCTUC Polo II - Pinhal de Marrocos\\3030-290 Coimbra, Portugal}
\authorinfo{Further author information: (Send correspondence to S.I.M.)\\S.I.M.: E-mail: stas@co.it.pt}

\begin{document} 
\maketitle

\begin{abstract}
  A theoretical framework for the quasi-monochromatic electromagnetic
  (EM) processes such as excitation and propagation of long wave
  packets in dispersive, dissipative, bianisotropic media with weak
  and slow nonlinearity is developed. The time-dependent EM fields
  associated with such processes are expressed as products of two
  functions: the slowly varying complex amplitude (SVCA) and the
  quickly oscillating carrier. The material parameters are treated as
  operators acting on time-dependent EM fields. By expanding these
  operators in the Maxwell equations in a series with respect to a
  small time scale parameter a system of equations for the SVCAs of
  the EM fields in such media is formulated. In the linear case, the
  dynamic equations for the SVCAs that correspond to the transverse
  components of the electric and magnetic fields resemble the vector
  transmission line equations. The obtained system of equations is
  used to derive the dyadic Green functions for the SVCAs of the EM
  fields in bianisotropic media. This framework is applied for
  modeling propagation of partially coherent EM radiation in a
  material whose parameters may depend on the amount of the EM energy
  that has passed through it. The same framework can be used in
  studying propagation of modulated EM waves through a waveguiding
  system that includes bianisotropic metamaterial
  components.
\end{abstract}


\section{\uppercase{Introduction}}
\label{intro}
In this paper we develop a theoretical framework for
quasi-monochromatic electromagnetic processes in dispersive media such
as propagation of long wave packets. The application of this approach
is the area of super-Planckian radiative heat transfer in
metamaterials~\cite{1,2,3,4,5,6}. In the processes that we consider,
most of the spectral energy is concentrated in a short interval around
a single frequency: $\o_0-\Delta\o/2\le\o\le\o_0+\Delta\o/2$,
$\Delta\o \ll \o_0$. Respectively, the time-dependent EM fields
associated with such processes can be expressed as products of two
functions: the slowly varying complex amplitude (SVCA) and the quickly
oscillating carrier. Because the characteristic time scales of the two
functions may differ by many orders of magnitude, the slowly
non-linear and dispersive response of the medium relates only to the
SVCA. Thus, the time-dependent electric and magnetic field vectors
$\_F \equiv \_E, \_H$ can be written in our case as
\[\_F(t) = {1\over
    2\pi}\int_{\o_0-\Delta\o/2}^{\o_0+\Delta\o/2}\_F_\o(\o)\,e^{-i\o
    t}d\o + \mbox{c.c.}={1\over 2}\,\_F_{\rm m}(t)\,e^{-i\o_0 t} +
  \mbox{c.c.}, \l{timedepf}
\]
where ``c.c.'' denotes the complex conjugate of the previous term. The
quantity
\[
  \_F_{\rm m}(t) = {1\over \pi}\int_{-\Delta\o/2}^{+\Delta\o/2}\_F_\o(\o_0+\O)\,e^{-i\O t}d\O
  \l{sva}
\]
is the SVCA (the time envelope) of the process. Note that for
monochromatic processes with delta-functional spectral density,
Eqs.~\r{timedepf} and \r{sva} imply that $\_F_{\rm m}$ is constant and
its magnitude simply equals the oscillating field amplitude.

From the above definition it follows that for any quantity
$\_G(t) = \=\alpha\.\_F(t)$, where $\=\alpha$ is a dyadic linear
integro-differential operator with the Fourier transform
$\=\alpha_\o= \=\alpha_\o|_{\o_0} +
(\o-\o_0)(\d\=\alpha_\o/\d\o)|_{\o_0} + \cdots$, the SVCA $\_G_{\rm m}(t)$ of the quantity $\_G(t)$ can be
expanded as
\[
  \_G_{\rm m}(t) = \=\alpha_{\o}|_{\o_0}\!\!\.\_F_{\rm
    m}(t)+i\left.{\d\=\alpha_{\o}\over\d\o}\right|_{\o_0}\!\!\!\!\.{\d\_F_{\rm
      m}(t)\over\d t} + \cdots,
  \l{expG}
\]
and the SVCA of the time derivative of the
same quantity can be expanded as
\[
  \left[{\d\_G(t)\over\d t}\right]_{\rm m} =
  -i\o_0\=\alpha_{\o}|_{\o_0}\!\!\.\_F_{\rm
    m}(t)+\left.{\d(\o\=\alpha_\o)\over\d\o}\right|_{\o_0}\!\!\!\!\.{\d\_F_{\rm
      m}(t)\over\d t} + \cdots.
  \l{expdG}
\]

In order to account for slow and weak nonlinearity, we assume that
$\=\alpha_\o$, which can be, \eg, dielectric permittivity dyadic at a
given point, is a function of a time-dependent parameter $T$ (\eg,
local temperature), which in turn depends on the amount of work the EM
field has performed on the material in the vicinity of the selected
point. For instance, if only the work due to the material loss
accounted by $\=\alpha_\o'' = \Im(\=\alpha)$\footnote{Here,
  $\Im(\overline{\overline{\alpha}}) =
  (\overline{\overline{\alpha}} -
  \overline{\overline{\alpha}}^\dagger)/(2i)$.} is taken into
account, then, at the selected point,
\[
T(t) = T(0) + {1\over 2 c_T}\int\limits_0^t\!d\tau\,\_F_m^\dagger(\tau)\.\=\alpha_\o''(T(\tau))\.\_F_m(\tau),
\l{localT}
\]
where $c_T$ is a parameter that has the meaning of the thermal
capacity of the material per unit volume. Such nonlinearity is
considered weak and slow when $c_T$ is enough large, so that
\[
  {1\over T}{dT\over dt} \ll {1\over |\_F_m|}{d|\_F_m|\over dt},
\]
which ensures applicability of expansions~\r{expG}
and~\r{expdG}. Therefore, when considering propagation of elementary
excitations (certain wave packets, more details ahead) with
characteristic duration $\tau_p \sim 2\pi/\Delta\o$, we may neglect
the change in $T$ (and, respectively, $\=\alpha_\o$) during time
intervals $\Delta t \sim \tau_p$, so that the material can be
considered effectively linear with respect to the propagation of such
a {\em single} pulse. However, propagation of a longer excitation
represented by a series of such pulses, will be influenced by the
gradual change in $T$ caused by every passing elementary
pulse. Thus, the slow and weak nonlinearity that we consider
works as an accumulative memory mechanism.

In what follows we derive equations which describe the dynamics of
such SVCAs of the electric and magnetic
fields in dispersive uniaxial bianisotropic media with such weak and
slow nonlinearity.

\section{\uppercase{Maxwell equations for slowly varying field amplitudes}}
In order to consider propagation of the elementary excitations
mentioned in the previous section, we consider the macroscopic Maxwell
equations in presence of both electric and magnetic
sources\footnote{The effective magnetic source current density $\_K$
  can be understood as resulting from the divergence-free, solenoidal
  electric currents.}, which can be written as
\[
  {\d\_D\over\d t} = \D\x\_H  - \_J ,\quad {\d\_B\over\d t} = -\D\x\_E - \_K,
\l{Maxwell}
\]
with the electric and the magnetic current densities $\_J$ and $\_K$
satisfying the continuity equations
\[
  \D\.\_J = -{\d\rho\over\d t},\quad \D\.\_K = -{\d\kappa\over\d t},
\]
where $\rho$ and $\kappa$ are the electric and magnetic charge
densities, respectively.

The electric displacement $\_D$ and the magnetic induction $\_B$ in
bianisotropic media satisfy the following constitutive relations,
written in a matrix form:
\[
\matr{\_D \\ \_B} = \matr{\=\E & \=\xi\\\=\zeta & \=\M}\.\matr{\_E \\ \_H} = \#M\.\matr{\_E \\ \_H}.
\l{matrel}
\]
In this relation, the components of the material matrix $\#M$ are
dyadic integro-differential operators\footnote{We understand these
  quantities as operators acting on the fields as functions of time,
  in order to take into account the material dispersion, and the
  parameter $T$, in order to account for the slow and weak
  nonlinearity.} corresponding to the constitutive parameters of the
medium: the effective permittivity $\=\E$, the effective permeability
$\=\M$, and the effective electro-magnetic and magneto-electric
coupling dyadics $\=\xi$ and $\=\zeta$. By substituting Eq.~\r{matrel}
into the Maxwell equations~\r{Maxwell} we obtain the
Maxwell equations in the so-called 6-vector form:
\[
  {\d\over\d t}\left(\,\#M\.\matr{\_E \\ \_H}\right) = \matr{0 & \D\x\=I \\ -\D\x\=I & 0}\.\matr{\_E \\ \_H}-\matr{\_J \\ \_K}.
  \l{matrMaxwell}
\]
In the frequency domain, the material relations~\r{matrel} can be written as
\[
  \matr{\_D_\o \\ \_B_\o} = \matr{\=\E_\o & \=\xi_\o\\\=\zeta_\o & \=\M_\o}\.\matr{\_E_\o \\ \_H_\o} = \#M_{\o}\.\matr{\_E_\o \\ \_H_\o}.
  \l{constrel}
\]
When interested in the quasi-monochromatic processes described by the
slow varying field amplitudes introduced in Sec.~\ref{intro}, we
can expand the Fourier-transformed material matrix $\#M_\o$ in a Taylor series in the
vicinity of the frequency $\o_0$:
\[
  \#M_{\o} = \#M_{\o}\big|_{\o_0} + \left.\O{\d\#M_{\o}\over\d\o}\right|_{\o_0} + \cdots,
  \l{Taylor}
\]
where $\O = \o - \o_0$, and drop the higher-order terms starting with
$\O^2$. Then, by using the results of Sec.~\ref{intro}, we find
that the slowly varying complex amplitudes $\_D_{\rm m}(t)$ and
$\_B_{\rm m}(t)$ satisfy
\[
  \matr{\_D_{\rm m} \\ \_B_{\rm m}} \approx
  \#M_\o\big|_{\o_0}\!\!\!\.\matr{\_E_{\rm m} \\ \_H_{\rm m}} +
  i\left.{\d\#M_\o\over\d\o}\right|_{\o_0}\!\!\!\!\.{\d\over\d
    t}\matr{\_E_{\rm m} \\ \_H_{\rm m}}.
\]
Therefore, the slowly varying amplitudes of the time derivatives of
the $\_D$ and $\_B$ vectors can be now written as
\[
  \left[{\d\over \d t}\matr{\_D \\ \_B}\right]_{\rm m} \approx
  -i\o_0\#M_\o\big|_{\o_0}\!\!\!\.\matr{\_E_{\rm m} \\ \_H_{\rm m}} +
  \left.{\d(\o\#M_\o)\over\d\o}\right|_{\o_0}\!\!\!\!\.{\d\over\d
    t}\matr{\_E_{\rm m} \\ \_H_{\rm m}}.  \l{dDdB}
\]
Note that the second derivatives of $\_E_{\rm m}$ and $\_H_{\rm m}$
with respect to time do not enter Eq.~\r{dDdB}, in agreement with
the accuracy of the expansion \r{Taylor}.

Based on the above results, we can now write the Maxwell equations for
the slowly varying amplitudes of the electromagnetic fields, by
replacing the corresponding terms in Eq.~\r{matrMaxwell} with their expressions in terms of the slowly varying
envelops. The result is
\[
  \left.{\d(\o\#M_\o)\over\d\o}\right|_{\o_0}\!\!\!\!\.{\d\over\d
    t}\matr{\_E_{\rm m} \\ \_H_{\rm m}} =
  \left(i\o_0\#M_\o\big|_{\o_0}+\matr{0 & \D\x\=I \\ -\D\x\=I &
      0}\right)\.\matr{\_E_{\rm m} \\ \_H_{\rm m}}-\matr{\_J_{\rm m}
    \\ \_K_{\rm m}}, \l{svaMax}
\]
where $\=I$ is the identity dyadic, and $\_J_{\rm m}$ and
$\_K_{\rm m}$ are the slowly varying amplitudes of the electric and
magnetic current densities, respectively.

It can be shown that in passive dispersive media with low dissipation,
the matrix $\d(\o\#M_\o)/\d\o$ is a positive definite matrix
and is, therefore, non-singular. In general, in all cases when the
mentioned matrix is non-singular we can rewrite Eq.~\r{svaMax} as
\[
  {\d\over\d t}\matr{\_E_{\rm m} \\ \_H_{\rm m}} =
  \left[{\d(\o\#M_\o)\over\d\o}\right]_{\o_0}^{-1}\!\!\!\!\.\left\{\left(i\o_0\#M_\o\big|_{\o_0}+\matr{0
        & \D\x\=I \\ -\D\x\=I & 0}\right)\.\matr{\_E_{\rm m} \\
      \_H_{\rm m}}-\matr{\_J_{\rm m} \\ \_K_{\rm m}}\right\}.
  \l{svaMax2}
\]
In what follows, we restrict our consideration only by such cases when
Eq.~\r{svaMax2} holds. Note that this form of Eq.~\r{svaMax2} in which
the time derivatives of the slowly varying amplitudes are expressed
through the spatial derivatives of the same quantities allows, for
instance, for a direct implementation of a numerical solver.

\section{\uppercase{Vector telegrapher's equation for wave packets in dispersive uniaxial bianisotropic media}}

An important special case of anisotropic media is when there exist a
single selected direction such that the electromagnetic response of
the medium along this direction is different from the response to the
field components orthogonal to the same direction. This is the case of
uniaxial bianisotropic materials, in which the constitutive parameters
assume the form
\[
  \=\E_\o = \=\E_{\rm t}+\E_{\rm n}\^z\ox\^z,\quad
  \=\M_\o = \=\M_{\rm t}+\M_{\rm n}\^z\ox\^z,\quad
  \=\xi_\o = \=\xi_{\rm t}+\xi_{\rm n}\^z\ox\^z,\quad
  \=\zeta_\o = \=\zeta_{\rm t}+\zeta_{\rm n}\^z\ox\^z,
  \l{uniax}
\]
where $\^z$ is the unit vector along the main anisotropy axis (which
we select coincident with the $z$-axis direction) and the subindex
$(\ldots)_{\rm t}$ denotes the dyadic components in the transverse
plane orthogonal to $\^z$. These dyadics satisfy
$(\^z\ox\^z)\xx\=\E_{\rm t}=\=\E_{\rm t}$,
$(\^z\ox\^z)\xx\=\M_{\rm t}=\=\M_{\rm t}$,
$(\^z\ox\^z)\xx\=\xi_{\rm t}=\=\xi_{\rm t}$,
$(\^z\ox\^z)\xx\=\zeta_{\rm t}=\=\zeta_{\rm t}$, where
$(\_a\ox\_b)\xx(\_c\ox\_d)\equiv(\_a\x\_c)\ox(\_b\x\_d)$.
In the reciprocal uniaxial bianisotropic media, the transverse dyadics
$\=\E_{\rm t}$ and $\=\M_{\rm t}$ reduce to
$\=\E_{\rm t} = \E_{\rm t}\=I_{\rm t}$,
$\=\M_{\rm t} = \M_{\rm t}\=I_{\rm t}$, where
$\=I_{\rm t} = \=I - \^z\ox\^z$, and the dyadics $\=\xi_\o$ and
$\=\zeta_\o$ are such that $\=\xi_\o=-\=\zeta_\o^{T}$, where
$(\ldots)^T$ denotes the dyadic transpose operation. In the general
uniaxial bianisotropic media, $\=\E_{\rm t}$ and $\=\M_{\rm t}$ may
have an additional antisymmetric component and $\=\xi_\o$ and
$\=\zeta_\o$ may be such that $\=\xi_\o \neq -\=\zeta_\o^{T}$.

Decomposition~\r{uniax} allows one to rewrite the Maxwell
equations in the uniaxial bianisotropic media in such a way that the
$z$-components of the fields are eliminated. Namely, when working in
the frequency domain, after some dyadic algebra the following result
can be obtained:
\[
  -i\o\,\tilde{\#M}_\o(\Dt)\.\matr{\_E_{\o,{\rm t}} \\ \_H_{\o,{\rm t}}} =
  \#J
  {\d\over\d z}\.\matr{\_E_{\o,{\rm t}} \\ \_H_{\o,{\rm t}}}-
  \#S_\o(\Dt)\.\matr{\_J_\o \\\_K_\o},
  \l{1dmaxwell}
\]
where $\Dt=\=I_{\rm t}\.\D$, $\_E_{\o,\rm t} = \=I_{\rm t}\.\_E_{\o}$,
$\_H_{\o,\rm t} = \=I_{\rm t}\.\_H_{\o}$,
$\#J = \matr{0 & \^z\x\=I_{\rm t}\\-\^z\x\=I_{\rm t} & 0}$, and
\[
  \tilde{\#M}_\o(\Dt) &= \matr{
    \=\E_{\rm t}    & \=\xi_{\rm t}\\
    \=\zeta_{\rm t} & \=\M_{\rm t}} +
  {1\over\o^2}\,
  \#J\.\Dt\matr{
    \E_{\rm n}    & \xi_{\rm n}\\
    \zeta_{\rm n} & \mu_{\rm n}}^{\!\!-1}\!\!\!\!\!\ox\Dt\.\#J,
  \l{tildeM}
\\
  \#S_\o(\Dt)\ &=
  \matr{
    \=I_{\rm t} & 0\\
    0          & \=I_{\rm t}}
  + {1\over i\o}\,\#J\.\Dt\matr{
    \E_{\rm n}    & \xi_{\rm n}\\
    \zeta_{\rm n} & \mu_{\rm n}}^{\!\!-1}\!\!\!\!\!\ox\^z.
  \l{bigS}
\]

The dynamic equations for the slowly varying amplitudes of the
corresponding quantities are immediately obtained from
Eq.~\r{1dmaxwell}, using the same procedure as for
Eqs.~\r{matrMaxwell} and \r{svaMax2}:
\[
  {\d\over\d t}\matr{\_E_{\rm m,t} \\ \_H_{\rm m,t}} =
  \left[{\d(\o\tilde{\#M}_\o(\Dt))\over\d\o}\right]_{\o_0}^{-1}\!\!\!\!\.
  \left\{\left(i\o_0\,\tilde{\#M}_\o(\Dt)\big|_{\o_0}+
  \#J{\d\over\d z}\right)\.
  \matr{\_E_{\rm m,t} \\ \_H_{\rm m,t}}-
  \matr{\tilde{\_J}_{\rm m,t} \\ \tilde{\_K}_{\rm m,t}}\right\},
  \l{sva1dmax}
\]
where
\[
  \matr{\tilde{\_J}_{\rm m,t} \\ \tilde{\_K}_{\rm m,t}} =
  \left(\#S_\o(\Dt) +
    i\left.{\d\#S_\o(\Dt)\over\d\o}\right|_{\o_0}\!\!\!{\d\over\d t}\right)\.
  \matr{\_J_{\rm m} \\ \_K_{\rm m}}.
\]
Eq.~\r{sva1dmax} can be understood as the Maxwell
equations~\r{svaMax2} reformulated as for an effectively
one-dimensional wave propagation along the $z$-axis. Indeed, in the
case when the material parameters are uniform in the transverse plane,
when taking a spatial Fourier transform of~\r{sva1dmax} in this plane the operator $\Dt$ transforms as
$\Dt\mapsto i\_k_t$, where $\_k_{\rm t}$ is the transverse wave
vector, and Eq.~\r{sva1dmax} immediately reduces to a system of two
differential equations in $\d/\d t$ and $\d/\d z$.

We can rewrite Eq.~\r{sva1dmax} slightly in order to highlight this
fact even further, using the fact that
$\#J\.\#J = \matr{\=I_{\rm t} & 0 \\ 0 & \=I_{\rm t}}$. After
multiplying Eq.~\r{sva1dmax} by $\#J$ and rearranging the terms we
find that
\[
  {\d\over\d z}\matr{\_E_{\rm m,t} \\ \_H_{\rm m,t}}
  = \left(i\#K(\o_0,\Dt)-{\d\#K(\o_0,\Dt)\over\d\o_0}\.{\d\over\d t}\right)\.
  \matr{\_E_{\rm m,t} \\ \_H_{\rm m,t}} + \#J\.\matr{\tilde{\_J}_{\rm m,t} \\ \tilde{\_K}_{\rm m,t}},
  \l{VTE}
\]
where $\#K(\o_0,\Dt) =
-\o_0\#J\.\tilde{\#M}_\o(\Dt)|_{\o_0}$. Eq.~\r{VTE} can be called the
vector telegrapher's equation for the SVCA of the electromagnetic field in bianisotropic media.

\section{\uppercase{Solution of vector telegrapher's equation: Envelop Green function}}
\label{homogen}

First, let us consider the situation when the source term in Eq.~\r{VTE}
vanishes. In this case, the vector telegrapher's equation reads
\[
  {\d\_F\over\d z} =
  \left(i\#K(\o_0,\Dt)-{\d\#K(\o_0,\Dt)\over\d\o_0}{\d\over\d
      t}\right)\.\_F, \l{VTE0}
\]
where $\_F = \matr{\_E_{\rm m,t} \\ \_H_{\rm m,t}}$. This equation can
be solved in the standard way, \eg~by separation of variables.

In order to find a more elegant way of resolving~\r{VTE0}, let us
consider a special case of Eq.~\r{VTE0} when the operators
$\#K(\o_0,\Dt)$ and ${\d\#K(\o_0,\Dt)/\d\o_0}$ commute\footnote{Which
  means that there exist a basis in which the two operators are
  diagonal.} and are independent of $z$. In this case the solution of
Eq.~\r{VTE0} can be formally written as
\[
  \_F(x,y,z,t) &=
      e^{z\left(i\#K(\o_0,\Dt)-{\d\#K(\o_0,\Dt)\over\d\o_0}{\d\over\d t}\right)}\.\_F_0(x,y,t)\nonumber\\
               &=
                 e^{iz\#K(\o_0,\Dt)}\.e^{-z{\d\#K(\o_0,\Dt)\over\d\o_0}{\d\over\d t}}\.\_F_0(x,y,t),
  \l{eqF0}
\]
where $\_F_0(x,y,t)$ is arbitrary. In particular, when
$\_F_0 = e^{-i\O t}\_A(x,y)$, where $\Omega\in[-\Delta\o/2;+\Delta\o/2]$ and
$\_A(x,y)$ is an eigenfunction of the operator $\#K(\o_0,\Dt)$ such
that $\#K(\o_0,\Dt)\.\_A = k_z \_A$, we obtain from Eq.~\r{eqF0}:
\[
  \_F_{\O,k_z} = e^{i k_z z}
  e^{-i\O\left(t-{\d k_z\over\d\o_0}z\right)}\_A(x,y),
\]
from which it follows that, in the considered case, the solutions of
Eq.~\r{VTE0} are wave packets with the polarization state and the
field distribution in the transverse plane described by the
eigenfunctions $\_A(x,y)$, the fast spatial phase variation described
by the term $e^{i k_z z}$, and with a slowly varying envelope function
$f(t-{\d k_z\over\d\o_0}z)$ with finite spectrum fitting into the
interval $[-\Delta\o/2;+\Delta\o/2]$. Such wave packets propagate
along the $z$-axis with the speed $v_{\rm g} = (\d k_z/\d\o_0)^{-1}$,
which is an analog of the group velocity in the considered medium.
Therefore, we can express the general solution $\_F(x,y,z,t)$ in this case as
\[
  \_F(x,y,z,t) = \sum_nf_n\!\left(t-z/v_g^{(n)}\right)e^{i k_z^{(n)} z}\_A^{\!(n)}(x,y),
  \l{gensol}
\]
where the summation is taken over all linearly independent
eigenfunctions $\_A^{\!(n)}(x,y)$.

Let us now consider the vector telegrapher's equation with a source term
under the same assumption that the operators $\#K(\o_0,\Dt)$ and
$\d\#K(\o_0,\Dt)/\d\o_0$ commute and do not depend on $z$. The
equation that we need to solve reads
\[
  {\d\_F\over\d z} =
  \left(i\#K(\o_0,\Dt)-{\d\#K(\o_0,\Dt)\over\d\o_0}{\d\over\d
      t}\right)\.\_F + \_S_0e^{-i\O t}\delta(z-z_0), \l{VTEin}
\]
where the source term is set to a slowly oscillating function
concentrated around the point $z = z_0$.
Knowing the solution of the homogeneous equation, we may look for a
particular solution of the inhomogeneous equation~\r{VTEin} in the
following form:
\[
  \_F_{s} &= \sum_{(+n)}C^{(+n)}\theta(z-z_0)e^{i k_z^{(+n)}z}e^{-i\O\left(t-{\d k_z^{\!(+n)}\over\d\o_0}z\right)}\_A^{\!(+n)}(x,y)\nonumber\\
  &+\sum_{(-n)}C^{(-n)}\theta(z_0-z)e^{i k_z^{(-n)}z}e^{-i\O\left(t-{\d k_z^{\!(-n)}\over\d\o_0}z\right)}\_A^{\!(-n)}(x,y),
\]
where $\theta(z)$ is the Heaviside step function, and the summations
over $(\pm n)$ mean the summation over the components with the
corresponding sign of the real part of the group delay factor
$\d k_z/\d\o_0$.

After substituting this into Eq.~\r{VTEin} and equating the
coefficients at the delta-functional terms we obtain
\[
  \sum_{(+n)}C^{(+n)}e^{i k_z^{(+n)}z_0}e^{i\O{\d k_z^{\!(+n)}\over\d\o_0}z_0}\_A^{\!(+n)}-
  \sum_{(-n)}C^{(-n)}e^{i k_z^{(-n)}z_0}e^{i\O{\d k_z^{\!(-n)}\over\d\o_0}z_0}\_A^{\!(-n)} = \_S_0.
\]

Without any loss of generality, we may assume that the eigenfunctions $\_A^{\!(n)}(x,y)\equiv|\_A^{\!(n)}\ket$ form a Riesz basis. Then, by introducing the reciprocal basis $\bra\tilde{\_A}^{(n)}|$ such that the inner product satisfies
\[
  \bra \tilde{\_A}^{(m)} | \_A^{\!(n)} \ket = \delta_{mn},
\]
where $\delta_{mn}$ is Kronecker's delta, we find that
\[
  C^{(\pm n)} = \pm e^{-i k_z^{(\pm n)}z_0}e^{-i\O{\d k_z^{\!(\pm n)}\over\d\o_0}z_0}\bra\tilde{\_A}^{\!(\pm n)}|\_S_0\ket.
\]
Respectively,
\[
  \_F_{s} &= \sum_{(+n)}\theta(z-z_0)e^{i k_z^{(+n)}(z-z_0)}e^{-i\O\left(t-{\d k_z^{\!(+n)}\over\d\o_0}(z-z_0)\right)}|\_A^{\!(+n)}\ket\bra\tilde{\_A}^{\!(+n)}|\_S_0\ket\nonumber\\
  &-\sum_{(-n)}\theta(z_0-z)e^{i k_z^{(-n)}(z-z_0)}e^{-i\O\left(t-{\d k_z^{\!(-n)}\over\d\o_0}(z-z_0)\right)}|\_A^{\!(-n)}\ket\bra\tilde{\_A}^{\!(-n)}|\_S_0\ket,
\]
from where, the dyadic Green function $\#G$ resulting from Eq.~\r{VTEin} can be expressed as
\[
  \#G(z,z_0,t) &= \sum_{(+n)}\theta(z-z_0)e^{i k_z^{(+n)}(z-z_0)}\tilde{\delta}\left(t-{\d k_z^{\!(+n)}\over\d\o_0}(z-z_0)\right)|\_A^{\!(+n)}\ket\bra\tilde{\_A}^{\!(+n)}|\nonumber\\
  &-\sum_{(-n)}\theta(z_0-z)e^{i
    k_z^{(-n)}(z-z_0)}\tilde{\delta}\left(t-{\d
    k_z^{\!(-n)}\over\d\o_0}(z-z_0)\right)|\_A^{\!(-n)}\ket\bra\tilde{\_A}^{\!(-n)}|,
    \l{green}
\]
where the function $\tilde{\delta}(t)$ must be physically understood as an
approximation of the Dirac delta function\footnote{Mathematically
  speaking, this function is $\tilde{\delta}(t) = (2\pi)^{-1}\int_{-\Delta\o/2}^{+\Delta\o/2}
  e^{-i\O t} d\O = (2\pi)^{-1}\Delta\o\,\mbox{sinc}(t\Delta\omega/2)$.}, concentrated on the
interval $-2\pi/\Delta\o\lesssim t\lesssim 2\pi/\Delta\o$.

Note that the obtained expression for the dyadic Green function~\r{green} can be
also formally written as
\[
  \#G(z,z_0,t) =
  e^{(z-z_0)\left(i\#K(\o_0,\Dt)-{\d\#K(\o_0,\Dt)\over\d\o_0}{\d\over\d
        t}\right)}\.\tilde{\delta}(t)\left(\theta(z-z_0)\#P^+ - \theta(z_0-z)\#P^-\right),
  \l{generalG}
\]
where
\[
  \#P^{\pm} = \sum_{(\pm n)}|\_A^{\!(\pm n)}\ket\bra\tilde{\_A}^{\!(\pm n)}|,
\]
are the projection operators that separate the waves with positive and
negative group delay factors. In fact, such a general form of this
result could be also obtained directly by considering the solution of
the homogeneous vector telegrapher's equation represented by
Eq.~\r{eqF0}. Therefore, Eq.~\r{generalG} with a suitable pair of
projection operators $\#P^+$ and $\#P^-$ can be considered as the
Green dyadic solution of~Eq.~\r{VTEin} even when the operators $\#K(\o_0,\Dt)$ and
${\d\#K(\o_0,\Dt)/\d\o_0}$ do not commute.

\section{Numerical results}

As a numerical example we consider propagation of a Gaussian wave
packet through a uniaxial reciprocal bianisotropic medium with the
transverse permittivity $\=\E_{\rm t}=\E_{\rm t}\=I_{\rm t}$, the
axial permittivity $\E_{\rm n}$, the transverse permeability
$\=\M_{\rm t} = \M_{\rm t}\=I_{\rm t}$, and the axial permeability
$\M_{\rm n}$. The electro-magnetic coupling dyadic
$\=\xi = -\=\zeta^T$ is such that
$\=\xi_{\rm t} = \xi_{\rm t}\^z\x\=I_{\rm t}$ and $\xi_{\rm n} =
0$. In this case, the propagating electromagnetic waves split into two
polarizations: TM polarization with $H_z = 0$ and TE polarization with
$E_z = 0$. Considering, for example, the waves of the TM polarization
propagating in the $xz$ plane, we can recast Eq.~\r{VTE0} in the basis
of such waves with the following substitutions
\[
  \_F = \matr{E_{{\rm m},x} \\ H_{{\rm m},y}}, \quad \#K = -\omega\#J\.\tilde{\#M}_{\o}, \quad
  \tilde{\#M}_{\omega} = \matr{\E_{\rm t} & -\xi_{\rm
      t}\\\xi_{\rm t} & \M_{\rm t}} + {1\over \o^2}\matr{0 & 0\\0 &
    \d_x\E_{\rm n}^{-1}\d_x}, \quad \#J = \matr{0&-1\\-1&0},
\]
where $\d_x = \d/\d x$.

\begin{figure}[b]
  \centering
  \includegraphics[width=0.4\linewidth,trim={0 0 0 3.5cm}]{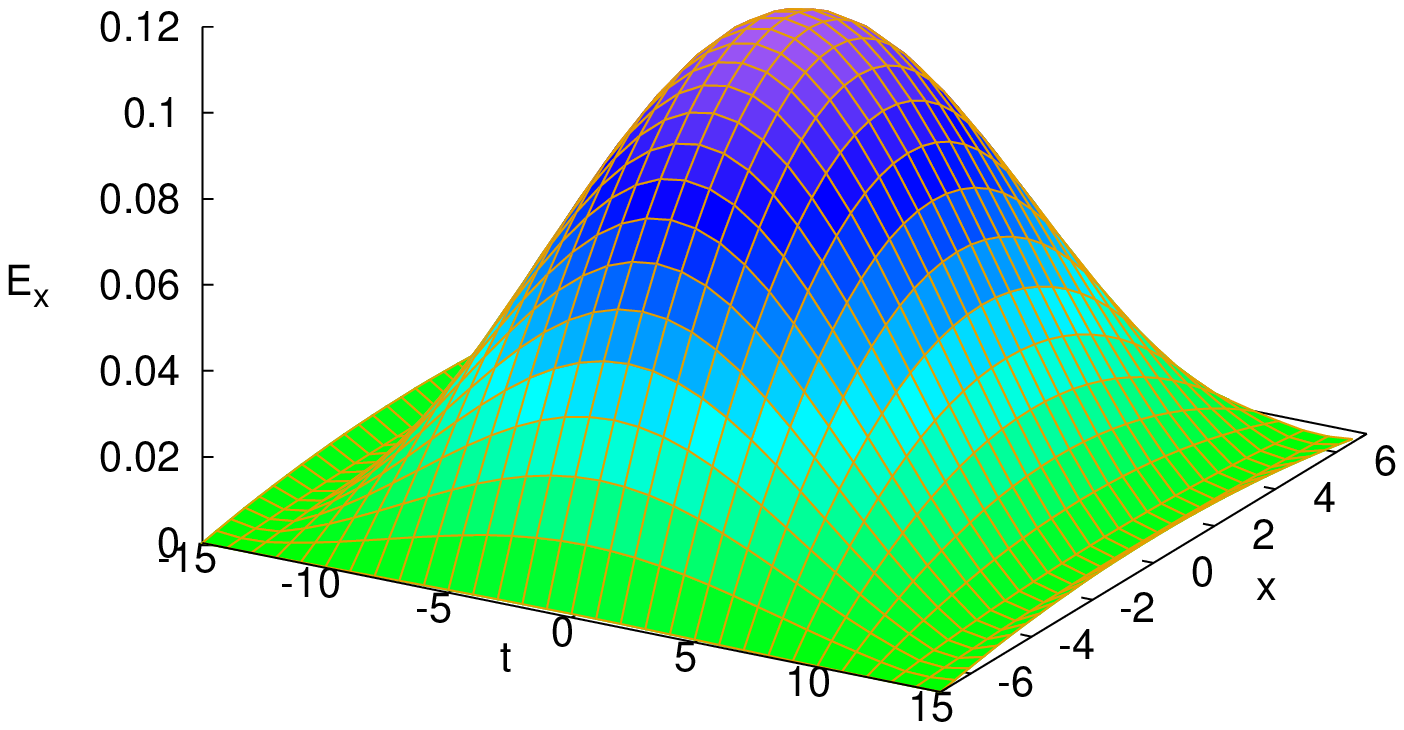}\hspace{1cm}
  \includegraphics[width=0.4\linewidth,trim={0 0 0 3.5cm}]{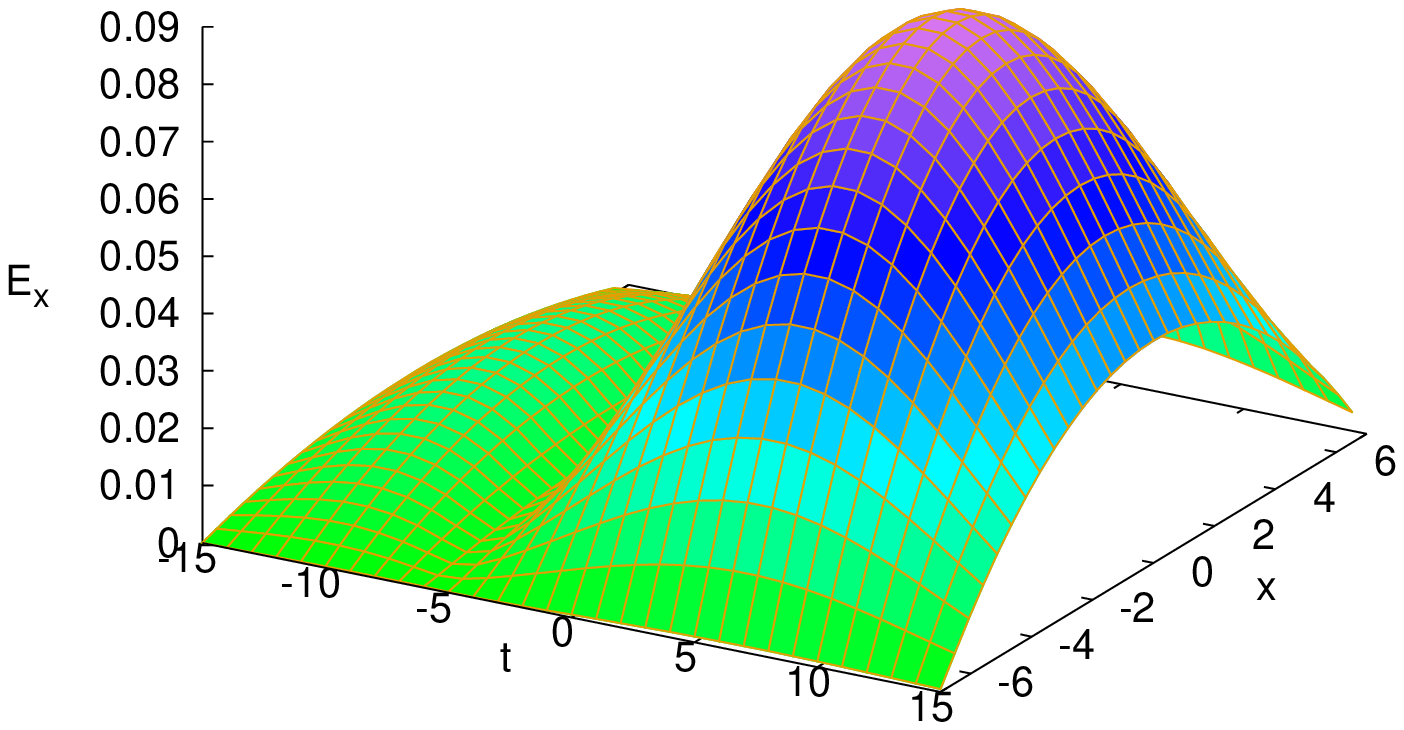}\\
  \includegraphics[width=0.4\linewidth]{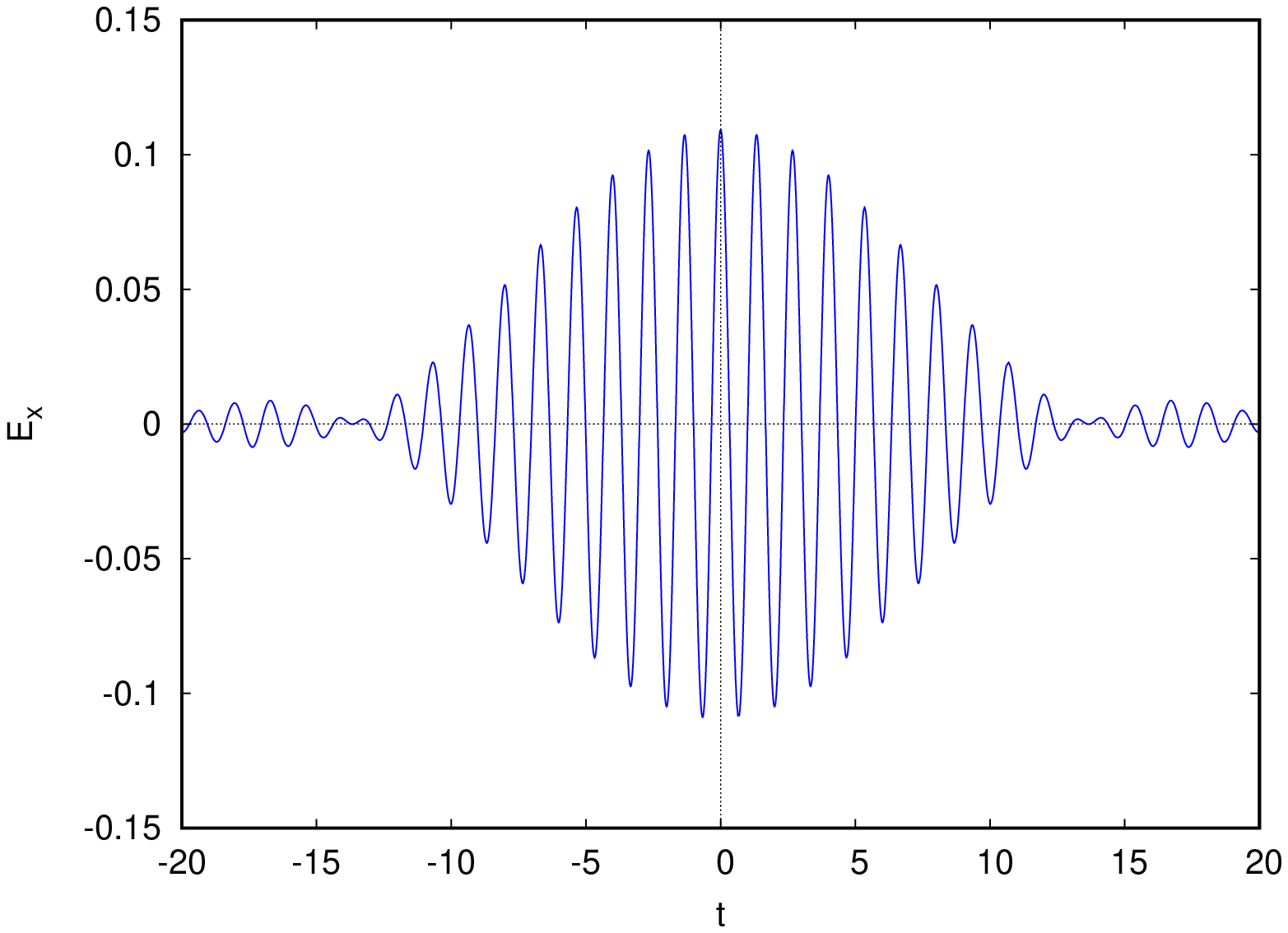}\hspace{1cm}
  \includegraphics[width=0.4\linewidth]{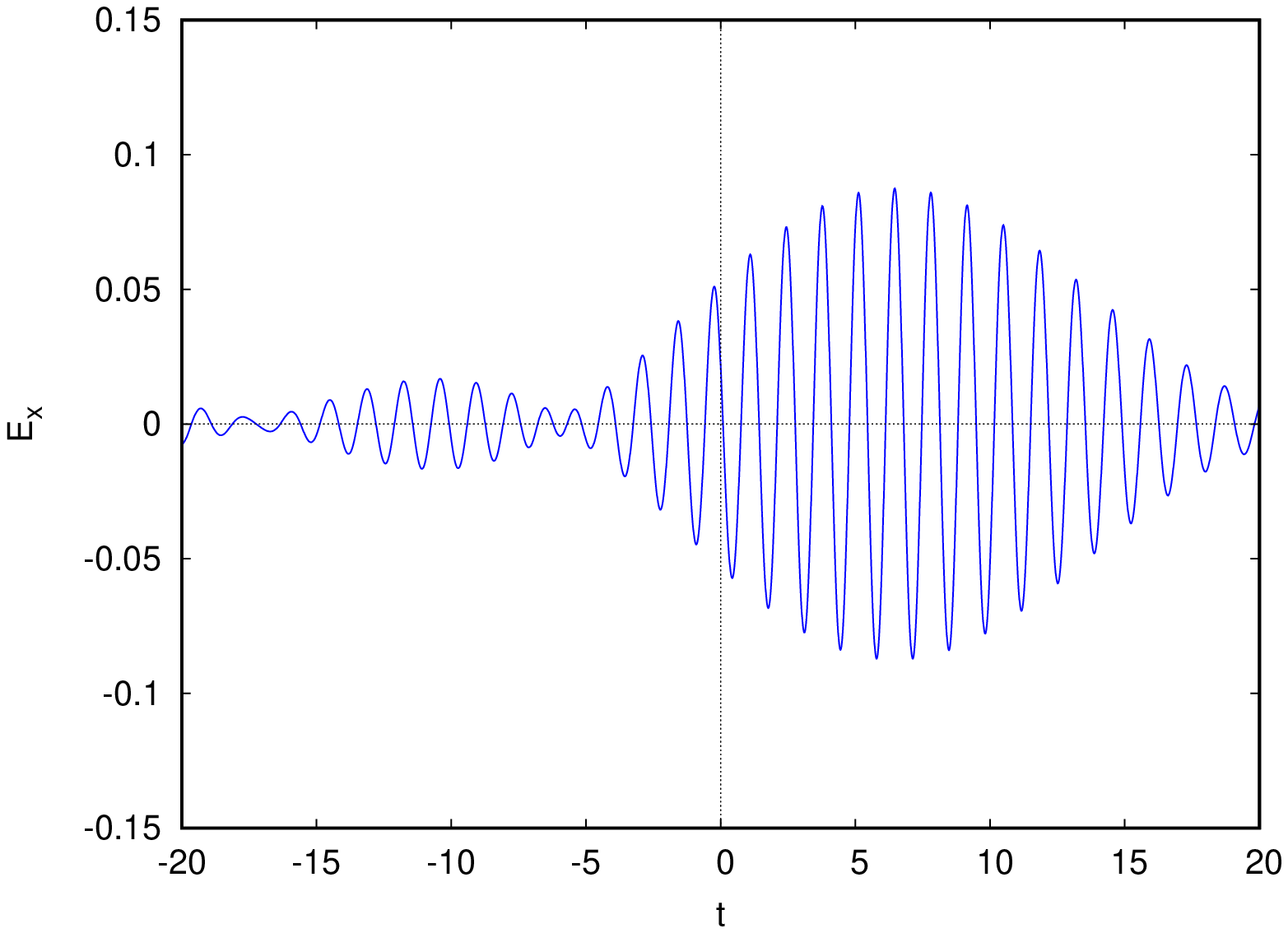}
  \caption{\label{fig1}Top: Envelope profiles (in $x$ and $t$) of a
    pulse propagating through a bianisotropic material with
    Lorentz-dispersive parameters (electric field, arbitrary units) at
    $z = 0$ (left) and $z = 100\,\Delta z$ (right). Bottom: Same as
    above at $x=0$ with fast oscillations included. The parameters are
    $\omega_0/\omega_{\rm p}=4.7$, $\o_{\rm res}/\o_{\rm p} = 5$,
    $\gamma/\omega_{\rm p}=0.1$, $\chi=0.3$,
    $\Delta z = 0.1c/\omega_0$. The units of $x$ and $t$ are such that
    $c = 1/\sqrt{\E_0\M_0} = 1$.}
\end{figure}
\begin{figure}[h!]
  \centering
  \includegraphics[width=0.47\linewidth,trim={0 0 0 3.5cm}]{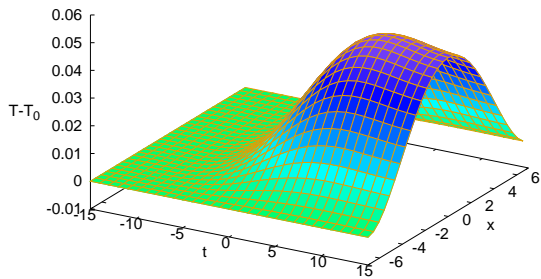}
  \caption{\label{fig2}Local temperature profile (in $x$ and $t$) of a
    pulse propagating through a bianisotropic material with
    Lorentz-dispersive parameters (arbitrary units) at
    $z = 50\,\Delta z$ and $c_T=1$. The other parameters are the same as in
    Fig.~\ref{fig1}.}
\end{figure}

In the case when the material parameters are not uniform (\eg~because
they depend on the local temperature $T$ which is not uniform), the
solution of such recast Eq.~\r{VTE0} can be attempted
numerically. We have implemented а symbolic calculation code that
allows one to evaluate $\_F(z)$ by numerical integration of
Eq.~\r{VTE0} (with 4th order Runge-Kutta method), from the initial
condition at $z=0$ to an arbitrary point $z=z_0$. The initial field at
$z=0$ is given in a form of a polynomial function in $x$ and $t$. When
$z$ increases, the evolution of $\_F$ in $x$ and $t$ is reflected in
the change of the coefficients of these polynomials. The numerical
integration routine also calculates the local temperature [Eq.~\r{localT}],
which may in turn affect the values of the material parameters.

In the following example, we consider a bianisotropic material with
Lorentz-type dispersion in $\E_{\rm t}$ and $\xi_{\rm t}$:
\[
  \E_{\rm t}(\o) = \E_0\left(1 + {\o_{\rm p}^2\over \o_{\rm res}^2 - \o^2
      - i\gamma\omega}\right), \quad
  \xi_{\rm t}(\o) = {i\sqrt{\E_0\M_0}\chi\o_{\rm p}^2\over \o_{\rm res}^2 - \o^2
      - i\gamma\omega},
\]
and $\E_{\rm n}=\E_0$, $\M_{\rm t}=\M_{\rm n}=\M_0$. The
characteristic evolution of the envelope of an (approximately)
Gaussian wave packet is shown in Fig.~\ref{fig1}. During the
propagation the pulse amplitude is decreased due to the loss and its
shape is distorted due to the dispersion. Fig.~\ref{fig2} shows the
associated $tx$-profile of the local temperature increase during the
propagation of the pulse.

\section{Conclusion}

In this article, we have developed a theoretical framework for
studying propagation of quasi-monochromatic signals through
dispersive, dissipative, bianisotropic media with weak and slow
nonlinearity. By representing the time-dependent EM fields associated
with such processes as products of the slowly varying complex
amplitude (SVCA) and the quickly oscillating carrier we have
formulated a system of equations for the SVCAs of the EM fields in
such media. In the linear case, these equations resemble the vector
transmission line equations, which can be solved analytically with the
Green function approach in the linear uniform case, or numerically in
other cases. This framework has been used to model propagation of wave
packets through a material whose parameters may depend on the local
temperature, which in turn is affected by the passing electromagnetic
radiation. The developed framework can be used for investigating the
propagation of modulated EM waves through waveguiding systems that
include bianisotropic metamaterial components and also in radiative
heat transfer-related problems involving such media.

%%%%%%%%%%%%%%%%%%%%%%%%%%%%%%%%%

%\appendix    %>>>> this command starts appendixes

\acknowledgments % equivalent to \section*{ACKNOWLEDGMENTS}

S.~I.~Maslovski acknowledges financial support from Funda\c{c}\~{a}o para a
Ci\h{e}ncia e a Tecnologia (FCT), Portugal, under Investigador FCT (2012)
grant (Ref. IF/01740/2012/CP0166/CT0002).

% References

%\bibliography{report} % bibliography data in report.bib
%\bibliographystyle{spiebib} % makes bibtex use spiebib.bst

\end{document}